\documentclass[]{pasj01}
\draft
\Received{}
\Accepted{}
 
\begin{document} 

\title{Spectrum composition of Galactic Center X-ray Emission with point and diffuse X-ray sources}
\author{
Masayoshi \textsc{Nobukawa}\altaffilmark{1},
and 
Katsuji \textsc{Koyama}\altaffilmark{2}}
\altaffiltext{1}{Faculty of Education, Nara University of Education, Takabatake-cho, Nara 630-8528, Japan}
\altaffiltext{2}{Department of Physics, Graduate School of Science, Kyoto University, \\
Kitashirakawa-oiwake-cho, Sakyo-ku, Kyoto 606-8502, Japan}
\email{nobukawa@cc.nara-edu.ac.jp}

\maketitle

\begin{abstract}

This paper reports that the X-ray spectrum from the Galactic Center X-ray Emission (GCXE) is expressed by the assembly of active binaries,  non-magnetic Cataclysmic Variables,  magnetic Cataclysmic Variables (X-ray active star: XAS), 
cold matter and diffuse sources. 
In the fitting of the limited components of the XASs, the GCXE spectrum exhibits
significant excesses with $\chi^2/d.o.f. =5.67$. The excesses are found at the energies of K$\alpha$, He$\alpha$, Ly$\alpha$ and radiative recombination continuum of S, Fe and Ni. By adding components of the cold matter and the diffuse sources, the GCXE spectrum is nicely reproduced with $\chi^2/d.o.f. = 1.53$, which is a first quantitative model for the origin of the GCXE spectrum. 
The drastic improvement is mainly due to the recombining plasmas in the diffuse sources, which indicate the presence of high-energy activity of Sgr A$^*$ in the past of $> 1000$~years.
\end{abstract} 

\section{Introduction}

The Galactic diffuse X-ray emission (GDXE) is excess X-rays from the Galactic plane over the uniform cosmic X-ray background (CXB) (\cite{Ko18} and reference therein). 
A major idea for the origin of the GDXE has been expected to be assembly of unresolved point sources (e.g. \cite{Re07}).
From the spectral similarity in the $\sim 2$--10~keV band, the candidate point sources have been listed as active binaries (AB) in low energy $\lesssim 2$~keV, and cataclysmic variables with magnetic white dwarfs (mCV) or non-magnetic white dwarfs (non-mCV). Throughout of this paper, these are called as the X-ray active stars (XAS).
Using the number densities and flux of the XASs in the solar vicinity, the origin of the GDXE spectrum has been proposed to be the assembly of the spectra of XASs
(e.g., \cite{No16, Ko18} and references therein).

The GDXE is spatially separated into the Galactic ridge X-ray emission (GRXE), Galactic bulge X-ray emission (GBXE), and the Galactic center X-ray emission (GCXE) (\cite{Ko18, Ya16} and references therein).  Their X-ray spectra are different with each others. 
The GCXE shows exceptionally strong line fluxes of heavy elements compared to those of the other regions, the GRXE and the GBXE.  
The origin of the GCXE has been proposed by many authors (e.g. \cite{No16, Xu16, Mu04}). 
\citet{Mu04} used the Chandra data from the region $\lesssim \timeform{9'}$ of Sgr A*, and suggested that the GCXE is composed by the XASs and some diffuse sources. 
\citet{Mu93} found that non-mCV has larger Fe abundances (equivalent widths) than those of mCV, hence proposed that the major component of XASs contributing the GDXE is the non-mCV. 

The bottle neck of the study of the GCXE origin is that high quality GCXE and XAS spectra have been limited, and samples for the XASs are also limited.
The high quality and statistics spectra have firstly come from the Suzaku satellite \citep{Mi07}. 
Then, the equivalent widths of Fe K-shall lines at 6.4, 6.7, and 6.97 keV ($EW6.4$, $EW6.7$, $EW6.97$) in the GCXE are found to be significantly larger than those of the GRXE and the GBXE (\cite{No16}).
\citet{No16, Ko18} arrived the same conclusion with \citet{Mu93} that the GCXE spectrum can be mainly composed of non-mCV.  However, the equivalent width of Fe K-shall line in the non-mCV is still smaller than those of the GCXE. This indicates that the GCXE needs new components with larger $EW6.4$, $EW6.7$, $EW6.97$ than any of the XASs. One possible source with large $EW6.4$ is an X-ray reflection nebula (XRN), which is a cold cloud irradiated by the past active Sgr A* \citep{Ry13}.  This corresponds cold mater (CM). To solve the $EW6.7$ and $EW6.97$ problem, \citet{Na13, No16} proposed diffuse sources of old-intermediate aged SNRs. 

This paper addresses a quantitative model of the GCXE spectrum, which introduces the diffuse sources and the CM in addition to the XASs. 
The contents are organized as follows.
Observation and data reduction are described in section 2. Section 3 is devoted for data analysis and results. In section 3.1, the X-ray spectra of the XASs are separately given. Section 3.2 shows the analysis method of the GCXE spectrum by the composition of the XASs using the best-fit spectra given in section 3.1.  In section 3.3, the goal of the GCXE model, composition of XASs, diffuse sources and CM, is described.  Section 4 is devoted for discussion and conclusion.

\section{Observation and Data Reduction} 

\begin{figure}[b] 
 \begin{center}
\includegraphics[width=16cm]{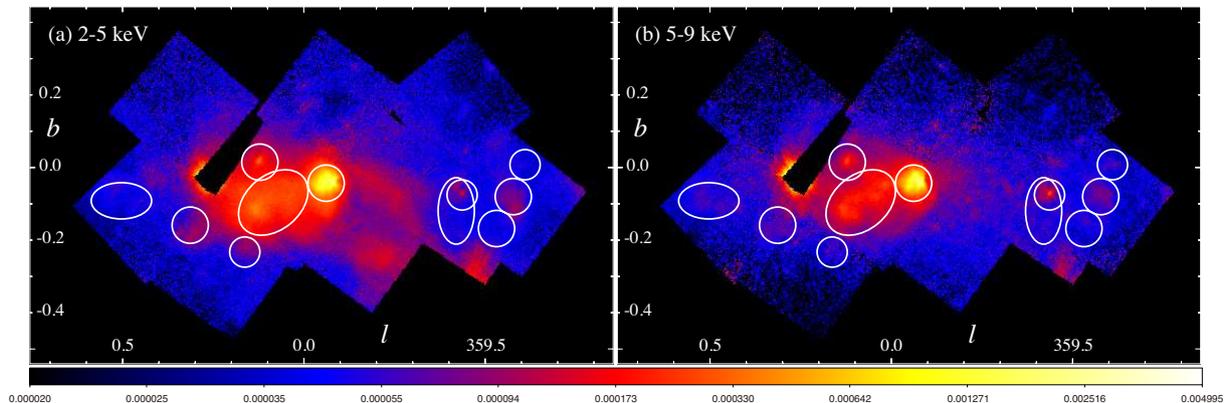} 
 \end{center}
\caption{Mosaic image of the GCXE observation in the 2--5 keV band (a) and 5--9 keV band (b).
The blank rectangle area is the region of bright point sources. The white circles indicate the diffuse X-ray sources.
}
\label{fig:image}
\end{figure}

We utilized the Suzaku archive data from the X-ray Imaging Spectrometer (XIS: \cite{Ko07a}) placed on the focal plane of the thin foil X-ray Telescopes (XRT: \cite{Se07}). 
The GCXE data are the sum of many pointing observations, with the total accumulation time of 1306~ks, given in table~\ref{tab:obs}.
Figure~\ref{fig:image} shows the region of the GCXE. The spectrum is made from the region within $(|l|, |b|) \lesssim (\timeform{0.6D}, \timeform{0.3D})$  excluding the white regions around the position of detected point or diffuse sources, including XRNs (\cite{Ko18} and reference therein). 
We used analysis tools in HEAsoft~6.28 and relevant CALDB released by the Suzaku team.


Although the energy resolution in the early observation was good enough \citep{Ko07b}, it becomes worse due to degradation of the charge transfer efficiency (CTE) in time. 
The CTE has been restored by the charge injection (CI) technique \citep{Uc09}. 
By the CI technique, the energy scale linearity is also restored.
However, the observation epochs are random and hence the energy resolution has degraded with increased line broadening. 
The resultant energy resolution is degraded by $\lesssim 50$~eV (FWHM) at 6--7~keV.
The calibration uncertainty of the energy scale linearity is $\sim 15$~eV in the energy band of the Fe and Ni lines\footnote{http://www.astro.isas.jaxa.jp/suzaku/doc/suzaku\_td/suzaku\_td.html}. 
To compensate these line broadenings and non-linearity effect, 
we applied {\tt Gsmooth} and {\tt gain} in the package of the spectral fitting code in XSPEC. 


\begin{table*}
  \tbl{List of the GCXE observations.}{
  \begin{tabular}{lcccc}
      \hline
Observation ID  & Position ($l$, $b$) 	& Start time & Stop time  	& Exposure (ks)$^*$  \\ 
      \hline
100027010 & (\timeform{0.039D}, \timeform{-0.088D}) & 2005-09-23 07:18:25 & 2005-09-24 11:05:19 & 44.8 \\
100027020 & (\timeform{359.736D}, \timeform{-0.059D}) & 2005-09-24 14:17:17 & 2005-09-25 17:27:19 & 42.8 \\
100037010 & (\timeform{359.737D}, \timeform{-0.059D}) & 2005-09-29 04:35:41 & 2005-09-30 04:29:19 & 43.7 \\
100037040 & (\timeform{0.038D}, \timeform{-0.087D}) & 2005-09-30 07:43:01 & 2005-10-01 06:21:24 & 43.0 \\
100048010 & (\timeform{0.037D}, \timeform{-0.086D}) & 2006-09-08 02:23:24 & 2006-09-09 09:06:15 & 63.0 \\
102013010 & (\timeform{0.037D}, \timeform{-0.087D}) & 2007-09-03 19:01:10 & 2007-09-05 05:20:20 & 51.4 \\
408017090 & (\timeform{359.944D}, \timeform{-0.045D}) &  2014-04-05 10:50:01 & 2014-04-05 23:38:12 & 22.2 \\
409011010  & (\timeform{359.943D}, \timeform{-0.049D}) & 2014-09-29 03:27:35 & 2014-09-29 16:00:15 & 20.2 \\
409011020 & (\timeform{359.946D}, \timeform{-0.050D}) & 2014-10-08 02:06:42 & 2014-10-08 11:05:09 & 17.3 \\
500005010 & (\timeform{0.445D}, \timeform{-0.099D}) & 2006-03-27 23:00:22 & 2006-03-29 18:12:15 & 88.4 \\
500018010 & (\timeform{359.451D}, \timeform{-0.077D}) & 2006-02-20 12:45:25 & 2006-02-23 10:50:14 & 106.9 \\
501008010 & (\timeform{359.825D}, \timeform{-0.205D}) & 2006-09-26 14:18:16 & 2006-09-29 21:25:14 & 129.6 \\
501009010 & (\timeform{359.906D}, \timeform{0.164D}) & 2006-09-29 21:26:07 & 2006-10-01 06:55:19 & 51.2 \\
502022010 & (\timeform{0.209D}, \timeform{-0.284D}) & 2007-08-31 12:33:33 & 2007-09-03 19:00:25 & 134.8 \\
503007010 & (\timeform{0.313D}, \timeform{0.151D}) & 2008-09-02 10:15:27 & 2008-09-03 22:52:24 & 52.2 \\
503072010 & (\timeform{359.588D}, \timeform{0.188D}) & 2009-03-06 02:39:11 & 2009-03-09 02:55:24 & 140.6 \\
505031010 & (\timeform{359.523D}, \timeform{-0.132D}) & 2010-09-25 12:36:56 & 2010-09-27 14:36:23 & 100.0 \\
508019010 & (\timeform{359.413D}, \timeform{-0.109D}) & 2013-09-24 06:22:37 & 2013-09-26 21:30:21 & 104.2 \\
508064010 & (\timeform{0.038D}, \timeform{-0.090D}) & 2013-09-20 10:45:46 & 2013-09-21 15:28:21 & 50.5 \\
      \hline
    \end{tabular}}\label{tab:obs}
\begin{tabnote}
\footnotemark[$*$] Effective exposure time.  \\ 
\end{tabnote}
\end{table*}

\begin{table} 
  \tbl{List of point sources.}{%
  \begin{tabular}{l}
\hline
Active Binary (AB) \\
\hline
\ \ GT~Mus, 
Algol, II~Peg, $\sigma$~Gem, UX~Ari, EV~Lac, HR~9024, $\beta$~Lyr, HD~130693 \\
\hline
Non-magnetic CV (non-mCV) \\
\hline
\ \ BF~Eri, BV~Cen, BZ~UMa, EK~TrA, FL~Psc, FS~Aur, GK~Per, KT~Per, SS~Aur, SS~Cyg, U Gem, V1159~Ori, \\
\ \  BV~Cen, BZ~UMa, EK~TrA, FL~Psc, FS~Aur, KT~Per, SS~Aur, SS~Cyg, U Gem, V1159~Ori, \\
\ \  V893~Sco, VW~Hyi, VY~Aqr, Z~Cam\\
\hline
Magnetic CV (mCV) \\
\hline      
\ \ Symbiotic Stars(SS): CH~Cyg, RS~Oph, RT~Cru, SS73-17, T~CrB,  V407~Cyg \\
\ \ Intemediate polar(IP): AO~Psc,	BG~Cmi, EX~Hya, FO~Aqr, GK~Per, IGR~J17195$-$4100, IGR~J17303$-$0601, MU~Cam,\\
\ \ \ \ \ \ NY~Lup, PQ~Gem, TV~Col, TX~Col, V1223~Sgr, 1RXS~J213344.1$+$51072, V2400~Oph, V709~Cas, XY~Ari \\
\ \ Polar(P): AM~Her, V1432~Aql, SWIFT~J2319.4$+$2619\\
\hline
    \end{tabular}}\label{tab:pointsource}
\begin{tabnote}
The source list is referred to \citet{No16} and \citet{Xu16}.  \\  
\end{tabnote}
\end{table}

\section{Analysis and Results} 

\subsection{X-ray spectra of the AB, non-mCV and mCV} 
Using the source list in table~\ref{tab:pointsource}, we made the flux weighted mean spectra from the individual sources of AB, non-mCV and mCV (XASs) by the same procedure as \citet{No16}.
These spectra were fitted by the model of {\tt Apec} + Fe~K$\alpha$ line.  In {\tt Apec}, the electron temperature $kT_{\rm e}$ and the abundance $z$ are free parameters. 
The energy bands for mCV, non-mCV and AB are 5--9~keV and 2--9~keV, respectively. 
For the Fe~K$\alpha$ line model, we used narrow Gaussian line at 6.4~keV. 

The best-fit spectral parameters of the XASs are listed in table~\ref{tab:tab3}. 
These values are for the mean spectra from each objects of XASs, not the mean of the best-fit values of each XASs (e.g. AB: \cite{Pa12}, non-mCV: \cite{By10}, mCV: \cite{Yu12}).
Since the best-fit value of $kT_{\rm e}$ for AB is biased to high flux sources with high $kT_{\rm e}$, higher temperature and flux than those of average, we limit the fit of $kT_{\rm e}$ to be bellow $\sim 1.1$~keV in the subsequent analysis.
The statistical errors in the former case are smaller than the variations of the source-to-source best-fit in the latter case (e.g. \cite{No16} and references therein).  

\begin{table}[h] 
\tbl{Best-fit parameters for AB, non-mCV, and mCV.}{
\begin{tabular}{lccc}
\hline
		& $kT{\rm e}$		& abundance ($z$)	& $EW6.4$ \\
		& (keV)			& (solar)			& (eV) \\ 
\hline
AB		& $3.3\pm{0.03}$	& $0.20\pm{0.01} $ 	&  43 \\
non-mCV 	& $7.4\pm{0.2}$	& $0.66\pm{0.03}$	& 100 \\
mCV		& $11.0\pm{0.5}$	& $0.25\pm{0.02}$	&  160  \\
\hline
\end{tabular}}
\label{tab:tab3}
\end{table}

\subsection{GCXE composed of XASs} 

Strong L$\alpha$ line of Au is found at the energy of 9.27 keV in the GCXE spectrum. This line is due to ionization of neutral Au (coated on the telescope surface) by cosmic-rays in the Suzaku orbit, and hence time variable \citep{Se07, Ta08}. Accordingly the GCXE spectrum near this energy has large ambiguity. We thus restrict the energy band to be 2--9~keV. 

Since the conventional idea for the origin of the GCXE is that the spectrum is consisted by the combination of the XASs, we apply the model of the GCXE spectrum by the sum of the spectral model for each XASs ({\tt Apec} + Fe~K$\alpha$ line). 
Here, we define the summed model as Model~A.
The temperature $kT_{\rm e}$ and abundance $z$ of XASs are free parameters. The $kT_{\rm e}$ and $z$ are searched within the variations in the observed results given by \citet{No16} in table 3 and 4.
The $N_{\rm H}$ values are free parameters. 
The spatial distribution of the AB would be uniform in the GC region, while those of the non-mCV and mCV would be more concentrated near the Galactic plane (near the center of GC). Therefore, the $N_{\rm H}$ values of non-mCV and mCV should be higher than that of AB. 

\begin{figure} 
 \begin{center}
\includegraphics[width=12cm]{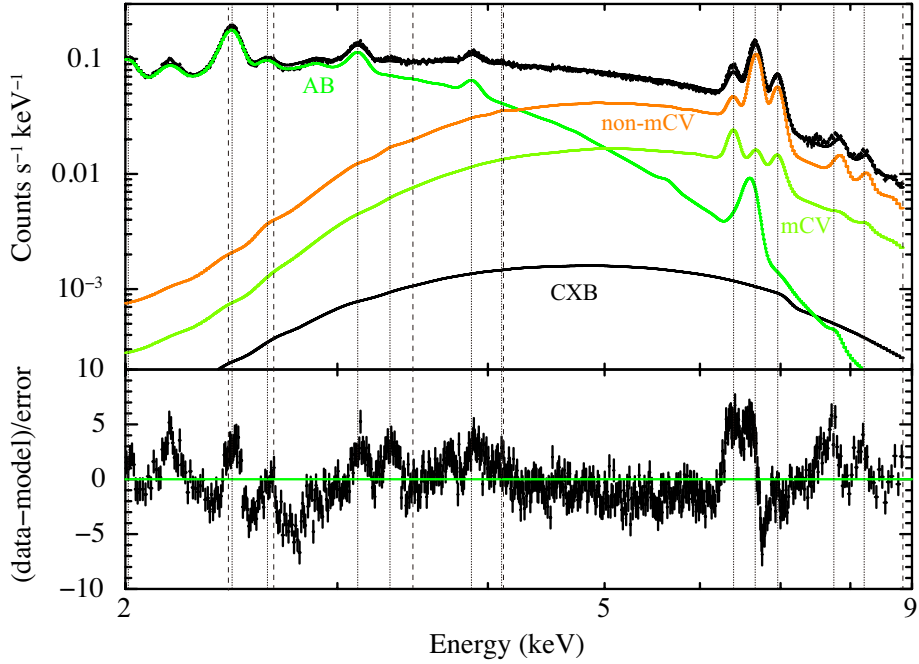} 
 \end{center}
\caption{Top: The GCXE spectrum with the combination of AB (green), non-mCV (orange) and mCV (yellow-green) spectra, here Model A. CXB is shown in black.
Bottom: residuals between the spectrum and the model.
The thin dotted lines indicate the energy of He$\alpha$ and Ly$\alpha$ of Si, S, Ar, Ca, Fe and Ni while the dashed ones shows their RRC energies in 2--9 keV.
}
\label{fig:modelA}
\end{figure}

The best-fit parameters and spectrum are given in table~\ref{tab:modelA} and figure~\ref{fig:modelA}.
There are significant residuals with large value of  $\chi^2/d.o.f.=3565/629$ (5.67).

\begin{table} 
 \tbl{Best-fit parameters for Model A, the composite of AB, non-mCV and mCV.}{
\begin{tabular}{lccccc}
      \hline
 Name 	& $kT_{\rm e}$ & abundance $(z)$ & $EW6.4$ & $N_{\rm H}$  & flux \\ 
 		& (keV)		& (solar) 			& (eV)		& ($10^{22}$~cm$^{-2}$) & \footnotemark[$*$]	  \\
      \hline
AB 		& $1.04\pm{0.01}$ 	& $0.48\pm{0.02}$ 	& 43		& $6.79\pm{0.03}$ 	& 9.0 \\     
non-mCV	& $7.4\pm{0.5}$	& $0.86\pm{0.20}$ 	& 100	& $22.1\pm{0.3}$ 	& 7.6 \\ 
mCV 	& $11.0\pm{2.1}$ 	& $0.25\pm{0.10}$ 	& 160	& $22.1\pm{0.3}$  	& 2.8 \\
      \hline
\end{tabular}}
\label{tab:modelA}
\begin{tabnote}
\footnotemark[$*$] Unit is $10^{-6}$~photons~s$^{-1}$~cm$^{-2}$~arcmin$^{-2}$.
\end{tabnote}
\end{table}

\subsection{GCXE composed of XASs and Diffuse sources} 

The result of Model A is incomplete. This is possibly due to large residuals at the energies of He$\alpha$ and Ly$\alpha$ and radiative recombination continuum (RRC) of Si, S, Fe and Ni. The position of the residuals are shown in figure~\ref{fig:modelA} by the dotted lines for He$\alpha$, Ly$\alpha$ and by the dashed lines for the RRCs. 
Then, we move on another idea that the origin of the GCXE is due to combination of the XASs, CM and diffuse sources (hereafter, Model B). 

\begin{figure} 
 \begin{center}
\includegraphics[width=12cm]{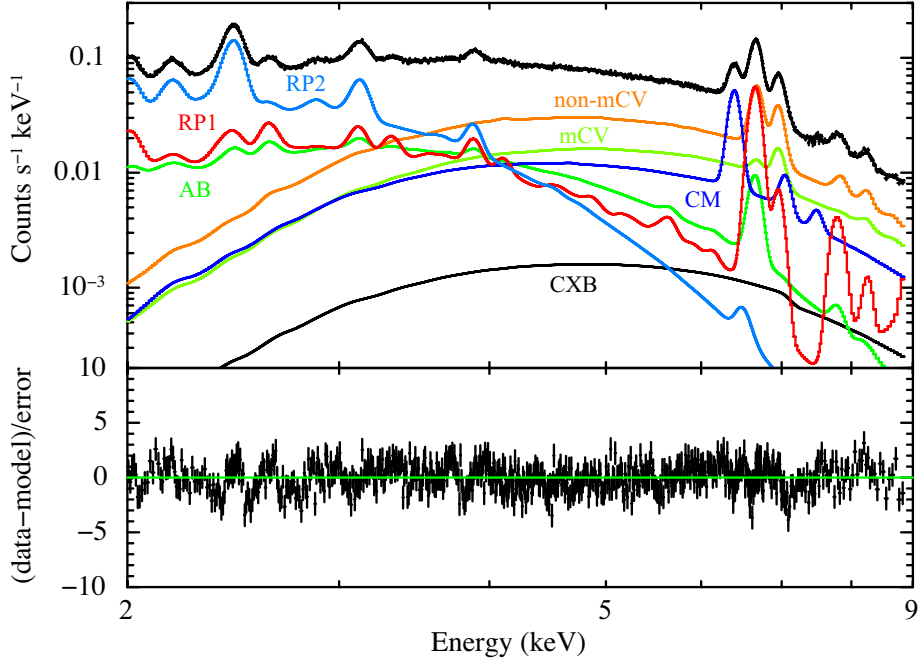}
 \end{center}
\caption{
Top: The GCXE spectrum with the combination of AB (green), non-mCV (orange), mCV (yellow-green), diffuse sources (RP1: red, RP2: cyan), and CM (blue) spectra (Model B). CXB is shown in black.
Bottom: residuals between the spectrum and the model.
}
\label{fig:modelB}
\end{figure}

The CM is a similar source as XRN, a giant molecular cloud irradiated by hard X-rays, but is different from the XRN in term of the size of the molecular cloud and irradiation source: not only hard X-rays but also high energy particles such as MeV protons. Thus, the CM has small scale and $N_{\rm H}$ to represent the excess of the Fe K$\alpha$ line. 

Since the main residual structures in Model A, He$\alpha$, Ly$\alpha$ and RRC would be due to recombining plasma (RP), we assume the spectra of the other diffuse sources in Model B as the RP ({\tt RNEI}). The spectrum of the diffuse source is made from some of candidate SNRs found with Suzaku, Chandra and XMM-Newton (e.g. \cite{Ts09, Na10, Po15}). The $kT_{\rm e}$ value is around $\sim 0.7$ keV. 
The initial temperature is fixed to be 10~keV. 
When ionization timescale $nt$ is searched in the range of $10^{10}$--$3\times 10^{13}$~cm$^{-3}$~s, two RPs (RP1 and RP2) with $nt\sim 3\times 10^{11}$~cm$^{-3}$~s and $nt\gtrsim 3\times 10^{12}$~cm$^{-3}$~s are found.
These two diffuse sources (CM and RP) would have very faint surface brightness, and hence has not been identified as the well  known diffuse sources.

The best-fit figure and parameters are given in figure~\ref{fig:modelB} and table~\ref{tab:modelB}.
The fit of Model B  is largely improved to  $\chi^2/d.o.f. = 931/608\, (1.53)$, far smaller than that of Model A, $\chi^2/d.o.f. =  3565/629 (5.67)$.
The reduced-$\chi^2$ value of 1.53 is not statistically acceptable.  
Since the statistics of GCXE spectrum are very high (total accumulation time is 1306 ks), non-statistical systematic error would be dominant in the GCXE fit. Therefore, we can safely regard that the fit of Model B is acceptable in the first order of approximation.

\begin{table*} 
\tbl{Best-fit parameter of AB, non-mCV, mCV, CM and Diffuse sources (RP1, RP2) in the GCXE.}{
\begin{tabular}{lccccc}
\hline \hline
\multicolumn{6}{c}{AB}  	\\ 
\hline
$kT_{\rm e}$\footnotemark[$*$]		&abundance ($z$)\footnotemark[$\dag$]	& ratio\footnotemark[$\ddag$]	 	& $EW 6.4$\footnotemark[$\S$]	& $N_{\rm H}$\footnotemark[$\|$] 	& flux\footnotemark[$\#$] \\
$1.8\pm{0.1}$ 		&$0.31\pm{0.01}$		&1.55			& 0.043			& $6.23\pm{0.01}$  	& 2.1 \\
\hline
\multicolumn{6}{c}{non-mCV} 		\\ 
\hline
$kT_{\rm e}$\footnotemark[$*$]	 	&abundance ($z$)\footnotemark[$\dag$] 	& ratio\footnotemark[$\ddag$]		& $EW 6.4$\footnotemark[$\S$]	& $N_{\rm H}$\footnotemark[$\|$]	& flux\footnotemark[$\#$] \\
$20\pm{1}$ 		&$0.85\pm{0.01}$		&1.29			& 0.10			& $13.4\pm{0.1}$ 	& 5.4 \\
\hline
\multicolumn{6}{c}{mCV} 	\\ 
\hline
$kT_{\rm e}$\footnotemark[$*$]	 	&abundance ($z$)\footnotemark[$\dag$]	& ratio\footnotemark[$\ddag$]	 	& $EW 6.4$\footnotemark[$\S$]	& $N_{\rm H}$\footnotemark[$\|$]	& flux\footnotemark[$\#$] \\
$8.6\pm{0.1}$ 		&$0.68\pm{0.04}$		& 2.72			& 0.16			& $13.4\pm{0.1}$ 	& 2.8 \\
\hline
\hline
\multicolumn{6}{c}{Cold Matter} 	\\ 
\hline
$\Gamma$		&				&				& $EW 6.4$\footnotemark[$\S$]	& $N_{\rm H}$\footnotemark[$\|$]	& flux\footnotemark[$\#$] \\
$2.1\pm{0.1}$ 		&				&				& $1.07\pm{0.01}$				& $13.4\pm{0.1}$ 	& 2.2 \\
\hline
\hline
\multicolumn{6}{c}{Diffuse source (RP1)} \\ 
\hline
$kT_{\rm e}$\footnotemark[$*$]		& $z$-L\footnotemark[$**$] 	&$z$-Fe\footnotemark[$**$] 	&$nt$\footnotemark[$\dag \dag$] 	& $N_{\rm H}$\footnotemark[$\|$]	& flux\footnotemark[$\#$] \\
$0.76\pm{0.03}$				& 1.3--1.8					& $0.9\pm0.1$				& $2.6 (\pm0.8)\times10^{11}$	& $6.23\pm{0.01}$ 	& 2.6 \\
\hline
\multicolumn{6}{c}{Diffuse source (RP2)} \\ 
\hline
$kT_{\rm e}$\footnotemark[$*$]		& $z$-L\footnotemark[$**$] 	&$z$-Fe\footnotemark[$**$] 	&$nt$\footnotemark[$\dag \dag$] 	& $N_{\rm H}$\footnotemark[$\|$]	& flux\footnotemark[$\#$] \\
$0.76\pm{0.03}$ 				&1.3--1.8					& $0.9\pm0.1$				&$ > 2.7\times10^{12}$ 	& $6.23\pm{0.01}$ 		& 4.2 \\
\hline
\end{tabular}}
\label{tab:modelB}
\begin{tabnote}
\footnotemark[$*$] Plasma (electron) temperature with unit of keV.
\\
\footnotemark[$\dag$] Abundance of the plasma with unit of solar. \\
\footnotemark[$\ddag$] The best-fit abundance relative to those in table 3. \\
\footnotemark[$\S$] Equivalent width of the Fe K$\alpha$ line. Unit is keV. \\
\footnotemark[$\|$] Absorption column density with the unit of $10^{22}$~cm$^{-2}$. \\
\footnotemark[$\#$] Unit is $10^{-6}$~photons~s$^{-1}$~cm$^{-2}$~arcmin$^{-2}$.  \\
\footnotemark[$**$] Abundance of light elements (Si--Ca) ($z$-L) and Fe ($z$-Fe) with unit of solar. \\
\footnotemark[$\dag \dag$] Ionization parameter ($nt$) with unit of cm$^{-3}$~s.
\end{tabnote}
\end{table*}

\section{Discussion and Conclusion} 

The Suzaku satellite is the first observatory which can provide the X-ray spectrum with high quality and statistics of the GCXE, and those of the XASs.  By the help of the Suzaku, Model A provides a quantitative test for the previous idea that GCXE is assembly of XASs (e.g. \cite{No16, Mu04} and references therein). As the result, Model A is rejected with large $\chi^2/d.o.f.= 5.67$. On the other hand, Model B includes extra ionization sources, aged SNRs and CMs with the Fe K$\alpha$ line. 

In the Model B fitting, we tried many RPs  for the diffuse sources in the wide range of $nt=10^{10} - 3\times 10^{13}$~cm$^{-3}$~s. These RPs have big clue to improve the fitting, particularly RP1 with the large RRC structure at the unique epoch of $nt = 2.6\times10^{11}$~cm$^{-3}$~s. The other RPs have all similar spectra with no large excess of the RRC structure (\cite{No16, Ko18}). As the result, the best-fit solution is converged to the two RPs (RP1 of  $nt = 2.6\times10^{11}$~cm$^{-3}$~s and RP2 of $nt \gtrsim 3\times10^{12}$~cm$^{-3}$~s). 
Thus, we find that Model B largely improves the fit to $\chi^2/d.o.f. =1.53$ due mainly to the recombining plasmas in the SNR spectra (RP SNRs). This would be the first quantitative judgment for the GCXE origin by XASs, and/or XASs plus diffuse sources and CM. 
The peculiar epoch at $nt \sim 3\times10^{11}$~cm$^{-3}$~s in RP1
leads to the notable effect in the $kT_{\rm e}\sim 0.8$ keV plasma to improve the fit in the line structures of Si, S, Ar, Ca, Fe and Ni (see the red-line in figure~\ref{fig:modelB}). 
The other plasma RP2 at $nt \gtrsim 3\times10^{12}$~cm$^{-3}$~s also has significant effect in the lines of S--Ca. 
Therefore, the RPs are additional tracers of the Sgr A* activity in the far past ($ > 10^3$~years) following the tracer of the XRN in giant molecular clouds at recent past of $ \lesssim 10^3$~years (e.g. \cite{In09, Ry13, Po15}). 

The abundances in this paper are those for the X-ray emitting plasma around the main star (either late type stars, or white dwarfs).  
In the table~\ref{tab:tab3}, the best-fit abundances of the XASs (the main star) are all smaller than 1 solar. This is unresolved mystery, but is beyond the scope of this paper.  
In the table~\ref{tab:modelB},  the abundance ratio relative to the lines in table 3 of AB (ratio$=1.55$) is larger than that of the non-mCV (ratio$=1.29$)\footnote{Contribution of the mCV to the GCXE flux is small, and the fraction is $\sim 0.3$ of that of non-mCV (see figure~\ref{fig:modelB}). }.
Since the spectrum of AB is dominated in the light element of S--Ca, but that of non-mCV is Fe. Thus the GCXE spectrum provide the first X-ray supports for the element dependent over-abundance in the GC region after the infrared observations \citep{Cu07, Ri07}.
We note that the abundance ratio in the non-mCV between table~\ref{tab:modelA} (Model A) and table~\ref{tab:tab3} is 2.4, far larger than Model B with the reasonable over-solar abundance. This is another support (other than $\chi^2$ value) that the conventional idea of Model A is quantitatively rejected.
 
The best-fit abundances in RP1 and RP2 are $\sim1.3$--$1.8$ solar for S--Ca, and $\sim 0.9$ solar for Fe. Thus the RP1 and RP2 would be the mixture of interstellar meduim near the GC ($\sim1.27$--$1.55$ solar) and the heavy element dominate ejecta of  SN origin.  The best-fit parameters in the faint CM of power-law index $\Gamma=2.1$ and $EW6.4=1$~keV can not constrain exactly the irradiation source, either MeV protons or hard X-rays above 7.1 keV (e.g. \cite{No15}). 

The best-fit $N_{\rm H}$ for the mCV,  non-mCV and CM are larger than those of the AB and the diffuse sources (RP1, RP2). 
This would be reasonable, if the spatial distributions of the mCV,  non-mCV and CM  are more concentrated toward the GC of larger $N_{\rm H}$ compared to the more uniformly (smaller $N_{\rm H}$ region) distributed of the AB and the diffuse sources (e.g. \cite{Na13}).


If the GRXE and GBXE data quality is significantly improved with future instruments, this enables us to study the origin of both the GRXE and the GBXE with the current accuracy of the GCXE study. Model B will provide a quantitative prediction for the origins of the GDXE (e.g., \cite{Ko18}). 
Elucidating the origin of the entire GDXE is our final goal with the Japanese next project, X-Ray Imaging and Spectroscopy Mission (XRISM; \cite{Ta20}).
 
\begin{ack}
The authors deeply appreciate the Suzaku team providing the high quality data.
This work is supported by JSPS KAKENHI Grant Numbers JP17K14289, JP20H01742, JP21K03615, and JP21H04493.
\end{ack}

\end{document}